\documentclass[aps,prl,twocolumn,showpacs,a4paper,superscriptaddress]{revtex4}
\usepackage{graphicx}
\usepackage{amsmath}
\newcommand{\degree}{\ensuremath{^\circ}}

\begin{document}
\title{Quantitative non contact dynamic Casimir force measurements}

\author{G. Jourdan}
\affiliation{Institut N\'eel, CNRS-UJF, BP 166 38042, Grenoble Cedex 9, France\\}
\affiliation{Universit\'e Joseph Fourier, BP 53 38041, Grenoble Cedex 9, France\\}
\affiliation{Laboratoire Kastler Brossel, CNRS-ENS-UPMC, 4 Place Jussieu, 75252 Cedex 05, France\\}

\author{A. Lambrecht}
\affiliation{Laboratoire Kastler Brossel, CNRS-ENS-UPMC, 4 Place Jussieu, 75252 Cedex 05, France\\}

\author{F. Comin}
\affiliation{ESRF, 6 rue Jules Horowitz, BP220, 38043 Grenoble Cedex, France\\}

\author{J. Chevrier}
\affiliation{Institut N\'eel, CNRS-UJF, BP 166 38042, Grenoble Cedex 9, France\\}
\affiliation{Universit\'e Joseph Fourier, BP 53 38041, Grenoble Cedex 9, France\\}
\affiliation{ESRF, 6 rue Jules Horowitz, BP220, 38043 Grenoble Cedex, France\\}

\date{\today}

\begin{abstract}
We show that the Casimir force gradient can be quantitatively measured with no contact involved. Results of the Casimir force measurement with systematic uncertainty of $3\%$ are presented for the distance range of 100-600 nm. The statistical uncertainty is shown to be due to the thermal fluctuations of the force probe. The corresponding signal to noise ratio equals unity at the distance of 600 nm. Direct contact between surfaces used in most previous studies to determine absolute distance separation is here precluded. Use of direct contact to identify the origin of distances is a severe limitation for studies of the Casimir forces on structured surfaces as it deteriorates irreversibly the studied surface and the probe. This force machine uses a dynamical method with an inserted gold sphere probe glued to a lever. The lever is mechanically excited at resonant frequency in front of a chosen sample. The absolute distance determination is achieved to be possible, without any direct probe/sample contact, using an electrostatic method associated to a real time correction of the mechanical drift. The positioning shift uncertainty is as low as 2 nm.
\end{abstract}


\maketitle

Quantum electromagnetic field fluctuations of the vacuum are the source of a quantum mechanical effect, the Casimir force, which is defined by the electron/photon coupling between two mirrors. Boundary conditions imposed on quantum electromagnetic field account for spatial dependence of this force \cite{Rodrigues}. Tailoring the mirrors shape and material may consequently result in an efficient way of monitoring this quantum phenomenon. In addition to a better understanding of the vacuum field fluctuations, study of the Casimir force aims at raising numerous issues related to MEMS/NEMS designs, since it was shown that the Casimir effect has a profound influence on the oscillatory behavior of such devices \cite{Chan}. Nowadays, experimental and theoretical works are more particularly concerned with thin film effects \cite{Lisanti,Lambrecht1} and optical properties of surface associated to materials \cite{Chen,Decca1}. In this context, nanostructured surface, metamaterials could also provide unusual force behavior as suggested by recent studies on plasmon surface polaritons~\cite{Intravaia,Henkel}. 

Since the effect of boundary on vacuum fluctuations is of primary importance, any experimental studies need to rely on a versatile instrument capable of accepting different surface samples with a defined probe. Quantitative measurements should be carried out avoiding any direct contact between a sphere and surface to prevent any irreversible damages to the surfaces. We call this a non-contact measurement. In this way one could ensure a reliable comparison between force curves measured at various surface points thanks to a XYZ positioning system that moves the sample stage over several millimeters. Thus, it becomes possible to compare directly the force curves measured for structured surfaces.
Beside lateral positioning, the force calibration and z positioning control issues turn out to be major limitations to be addressed in order to carry out experimental program.

\begin{figure}[b]
	\centering
		\includegraphics[width=8cm]{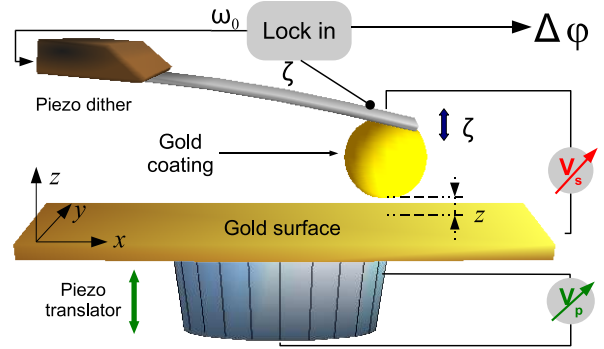}
	\caption{Scheme of the experimental setup up.	The XYZ positioning system attocube, on which the sample stage is mounted and the optical fiber based interferometer above the end of the microlever are not shown for simplicity.}
	\label{fig:figure_1}
\end{figure}

As shown in Fig.~\ref{fig:figure_1}, a microsphere with a radius $R_s$ of about 20~$\mu$m is glued to the end of an AFM microlever (NSC 18 MikroMasch), thus forming the Casimir microscope force probe. The overall probe is then coated with a 30~nm thick titanium layer followed by a 300~nm thick gold layer. The Casimir effect in the present experiment arises between the bottom part of the microsphere and a flat gold surface. Roughness of the two interacting surfaces are respectively lower than 3 and 2~nm rms as measured by AFM. AFM image analysis of the microsphere have also shown that the radius dispersion is smaller than $\pm 20$~nm over a $25 \; \mu \mathrm{m}^2$ cap, which is relevant for the Casimir study. 
The probe turns the force signal into lever motion detected with an optical fiber based interferometer.

\begin{figure}[t]
	\centering
		\includegraphics[width=8cm]{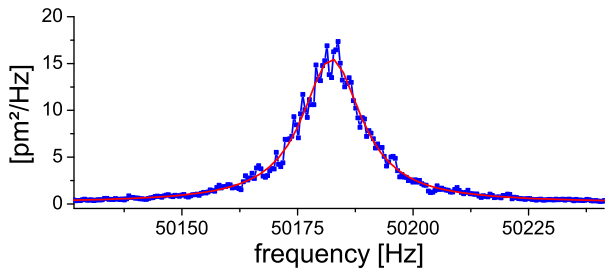}
	\caption{Noise spectrum density of the microlever oscillations around its resonance frequency. The Brownian motion peak exhibits a Lorentzian shape profile that is consistent with an harmonic oscillator model. Fit parameters characterize the mechanical response as follows : $f_0 = 50182.4$ Hz, $\gamma = 98$ rad.s$^{-1}$, position standard deviation generated by thermal bath coupling $< \zeta^2 > = 376 \pm 30$ pm$^2$. The detection noise at 0.11 pm$^2/$Hz is negligible in vicinity of the resonance frequency. Force noise analysis sets the force resolution at 13~fN/$\sqrt{\mathrm{Hz}}$, when working in dynamic mode. As a result, the smallest force gradient that can be detected is 1.3~$10^{-6}$~N.m$^{-1}$ when the oscillation amplitude $a_0$ is set to 10.2 nm. }
	\label{fig:figure_2}
\end{figure}

In this experiment the force probe, that can be considered as an harmonic oscillator, is mechanically excited at its free resonance frequency $\omega_0 = 2 \pi \times 50182$ rad.s$^{-1}$ with an amplitude $a_0=10.2$ nm measured at the far end of the lever. The Casimir force exhibits a spatial dependence $F_C(z)$ that modifies the natural stiffness $k_0$ of the oscillator by the force gradient $F'_C(z)$ in the linear regime $k_{\mathrm{eff}} = k_0  - F'_C$ \cite{Decca1,Chan2}. A lock-in demodulates the motion signal at the frequency $\omega_0$ and provides its phase change $\Delta \phi$ disturbed by the force gradient: 
\begin{equation}
\tan \Delta \phi = \frac{\omega_0}{\gamma} \frac{1}{k_0} F'
\label{eq:phase_shift1}
\end{equation}
The damping rate of the oscillator related to the friction coefficient $\Gamma = \gamma/m$ is measured to be $\gamma = 98$ rad.s$^{-1}$ (Fig.~\ref{fig:figure_2}). This method is simple, its main advantage consists in the fact that gradient measurement is not affected by dither piezo response.

Force calibration and piezo extension controls are the key points that determine the quality of the here presented measurements. Sample positioning precision and stability at the level required for the Casimir force measurement in the open loop are controlled by quality of z piezo calibration and affected by mechanical drift.
The sphere-plate motion can actually be broken down into two components. First, the piezo extension can be accurately enough controlled during a force approach-retract curve. A triangular shape voltage $V_p$ with an amplitude of 10 V at a frequency of 2 mHz is continuously applied on the fine piezoelectric translator, the z piezo. Displacements are calibrated using the optical fiber based interferometer. The calibration is then valid only for these 3 parameters: triangular shape, the maximum extension of 10 V, the scan speed of 2 mHz. Non-linear fits enable us to account for the hysteresis between the approach and withdrawal motions. The second point is the drift with the usual range of 1 nm.min$^{-1}$. It is related to the mechanical stability. This undesirable motion could originate from the thermal instability of the set up. Change in temperature of the overall frame as small as 0.01 K.min$^{-1}$ can account for this drift for the standard dilatation coefficient of $\alpha \approx 10^{-6}$ K$^{-1}$ and a mechanical loop length of $\approx 10$ cm. In the presence of this drift the Non-Contact requirement raises a major challenge in determining the absolute separation distance. However mechanical drift rate appears to be constant in time during a few scans and we have been able to correct it safely when processing data. In this way we eliminated the drift effects (dilatation and contraction) during the cycle phases.

Force gradient measurements and the absolute distance determination are ensured by the use of the electrostatic force. This can be safely done as its mechanical stresses on the sphere are geometrically identical to the one associated to Casimir force. 
Applying a bias voltage $V_s$ between the microsphere and the flat surface sets up an attractive force.
Its gradient can be described by sphere-plate capacitance second order derivative $C''$, the residual potential $V_0$ associated to the two surfaces \cite{Chan,Decca1} and is given by $F'_e=\frac{1}{2} C''(z) (V_s - V_0)^2$. Although an exact expression of $C''$ could be used \cite{Durand}, using its asymptotic form appears more convenient and precise enough:
\begin{equation}
F'_e = \frac{\pi \epsilon_0 R_s}{z^2} (V_s-V_0)^2
\label{eq:electrostatic_force_gradient}
\end{equation}
The discrepancy are estimated to be lower than $1\%$ for the distances $z<500$ nm and the radius $R_s=20$ $\mu$m. The Casimir force is evaluated within the framework of the proximity force approximation \cite{Derjaguin, Bordag}. As a result, gradient force measurement can  be compared directly to the Casimir pressure model between two parallel plates \cite{Decca1}: $P_{//} = F'_C(z)/ (2 \pi R_s)$. As in relation~(\ref{eq:electrostatic_force_gradient}) the force gradient is proportional to $R_s$, it turns out to be more relevant to calibrate the probe through sensitivity $\beta$:
\begin{equation}
\frac{F'_e}{2 \pi R_s}= \beta \tan \Delta \phi
\label{eq:calibration_sensitivity}
\end{equation}
The sphere radius $R_s$ is actually not required for the experiment to theory comparison.

The residual potential $V_0$ is found to be equal to $75 \pm 3$~mV and it remains stable below the distance of 1 $\mu$m during experimental sessions.
 When sweeping simultaneously the bias voltage $V_s \in [-1;1]$ V at 20 mHz and the sphere-plate distance in the 100-600 nm separation range (conditions are described above), the force gradient surface $P=\tan[ \Delta \phi(V_s,V_p)]$ is mapped and includes both forces: the electrostatic force superimposed to the Casimir force. For better accuracy during electrostatic calibration, phase shifts greater than $15\degree$ are removed, since small error in phase origin generates an large error for evaluating $\tan \Delta \phi$ that could be prejudicial for $\beta$ assessment.
Dilatation of $z$ axis can then be taken into account with a mechanical drift of 1.0 nm.min$^{-1}$ estimated by comparing two successive approach-retract cycles : this correction turns out to automatically ensure a good agreement between values of $\beta$ coefficient evaluated for the retract and the withdrawal. Without any corrections, the drift effect takes away the values of the fit parameters, since it produces a contraction and an extension or vice versa of the runs associated to these two successive phases.
In order to evaluate $\beta$, Casimir component obtained by interpolating selected data with $V_s = V_0$ is subtracted over the total surface. This curve is called $P_C$ and will be used below. In this way, the resulting electrostatic component $P_e(z,V_s)$ can be fitted using Eqs.~(\ref{eq:electrostatic_force_gradient}) and~(\ref{eq:calibration_sensitivity}), thus providing $\beta = 27.9 \pm 0.3$~N.m$^{2}$ and $z_0 = 588.5 \pm 2$ nm the position of contact.

At that stage using Eqs.~(\ref{eq:phase_shift1}), (\ref{eq:electrostatic_force_gradient}) and (\ref{eq:calibration_sensitivity}), the lever stiffness can be written as $k_0 = 2 \pi R_s \omega_0 \beta / \gamma$. Therefore $k_0$ is estimated to 11.3 N.m$^{-1}$ and is consistent with the value 11.8 N.m$^{-1}$ evaluated within the equipartition theorem $\frac{1}{2}k_{\mathrm{B}}T = \frac{1}{2}k'<\zeta^2>$, where $k_0 \cos^2 \theta = k'$ takes into account the $15\degree$ tilt of the lever with respect to the surface. The uncertainty of evaluating $k_0$ using this method is partially related to the lever motion detection position, ie the optical fiber position. The stiffness is larger than indicated by the manufacturer ($3.5 \pm 2$ N.m$^{-1}$) and can be explained by the metallic coating on both sides of the lever and by the position of the sphere center 30 $\mu$m away from the end of the microlever. We have checked that cantilever static deflection generated by Casimir force is negligible in the studied separation range ($z>100$ nm).

The Casimir force measurement then consists in setting up the bias voltage $V_s$ equal to the residual potential $V_0$ and in carrying out sphere-plate distance sweeping according to the previous piezo extension cycle. Great attention has again been paid to measuring successively several cycles and to correcting for mechanical drift when processing data.
 After scaling the force gradient using $\beta$, overlapping the new data sets with the Casimir curve reference $P_C$ obtained during the previous calibration enables us to determine the absolute position within $\pm 0.5$ nm.
 
Finally two significant issues have still to be raised before performing a close comparison with theory. First, drift in the oscillator resonance frequency is also observed and can be explained by change in temperature mainly through its Young modulus thermal sensitivity. For silicon cantilever sensitivity is found to be $(\partial f_{res}/ \partial T)/f_{res} = -5.2 \; 10^{-5}$ K$^{-1}$ \cite{Cleland,Giessibl}. Resulting phase drift appears also to be constant for a few scan cycles and is evaluated to $3.1 \; 10^{-6}$ rad.s$^{-1}$ when comparing successive curves at long distance where force gradient is not sensitive to mechanical drift. A thermal drift of only $10^{-3}$ K.min$^{-1}$ can account for this observed rate and is consistent with the one estimated before for mechanical drift.
Error generated over one cycle period $\delta \phi_d = 7.9 \; 10^{-4}$ rad ($T/2=250$ s) is larger than the phase thermal noise $\delta \phi_n = 2.3 \; 10^{-4}$ rad.
We applied a subtraction to this drift in order to perform weak force gradient measurement at long range. Secondly, a 500 ms constant time low pass is implemented at the output of the lock-in in order to improve gradient force resolution. Given the scanning velocity $v=1.9$ nm.s$^{-1}$, it results in averaging gradient force curve over 1~nm, which can therefore disturb sharp variation measurement.
 Signal filtering requires slow sweeping and consequently high set up stability: in this context drifts appear as the major limitations for this experiment. Here the filtering effect proved to be negligible for the chosen experimental parameters since the approach-retract curves independently processed completely overlap. Moreover it means that drifts and hysteresis corrections are efficient and consistent.

\begin{figure}
	\centering
		\includegraphics[width=8cm]{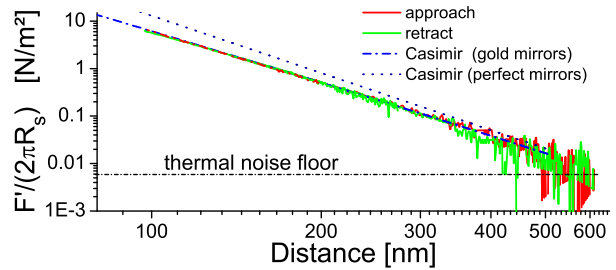}
	\caption{The Casimir force gradient measurement in the 100-600 nm range. The graph displays experimental data sets associated to the retract and approach phases as lines. The dashed line corresponds to the Casimir force theoretical model that takes into account the finite conductivity of gold, whereas the dotted line indicates the perfect behavior of the force derived in 1948 by Casimir himself.}
	\label{fig:figure_3}
\end{figure}

\begin{figure}
	\centering
		\includegraphics[width=8cm]{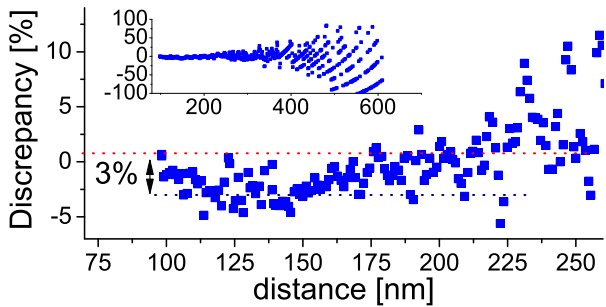}
	\caption{Discrepancy between the force gradient measurement and the theoretical model. At short distance, systematic uncertainty dominates over the experimental statistical uncertainty and the theoretical assessment uncertainty. Here a $3 \%$ deviation is observed with a specific theoretical curve, therefore it is consistent with the experimental error analysis.}
	\label{fig:figure_4}
\end{figure}

Fig.~\ref{fig:figure_3} displays two gradient force measurements $F'/(2 \pi R_s)$ in pressure unit [Pa] associated to the approach and the retract phases. The dashed line corresponds to the Casimir pressure between two parallel plates computed within the framework of quantum optical scattering theory using only optical data as free parameters to decsribe the material properties \cite{Genet}. Recent computations \cite{Pirozhenko} have emphasised sample dependence on evaluating theoretical Casimir pressure $P_{//}$. For gold mirrors the intrinsic uncertainty has been estimated of the order of $5 \%$ and therefore it limits the present experiment to theory comparison. Nevertheless, our experiment demonstrates again the finite conductivity effects on real Casimir force, which can be compared with the ideal one in dotted line in the graph. Regarding the PFA validity, which is the second major concern in the theoretical assessment, recent studies \cite{Emig} suggested that a discrepancy should be smaller than $1 \%$ in our configuration at least at distances below $z<200$ nm. In this context, as shown in Fig.~\ref{fig:figure_4}, the experiment to theory comparison does not reveal any disagreement, since the discrepancy is around $3 \%$ at the shortest distance for this particular theoretical curve.

Actually systematic uncertainties dominate over theoretical assessment as well as experimental data. This is particularly true in the second case at short distance. Shift in the distance origin as small as 1.2 nm or a drop by $3 \%$ of $\beta$ can indeed turn the systematic discrepancy in Fig.~\ref{fig:figure_4} into random discrepancy.
It shows that previous experimental error analysis is consistent in assessing $\beta$ and $z_0$ respectively at a few percent level (electrostatic model and calibration uncertainty) and at $\pm 2.1$ nm ($\pm 2$ nm for the Casimir reference curve and $\pm 0.5$ nm for second positioning). Further experiments with this set up have reduced this uncertainty down to about $\pm 1$ nm.

As shown in Fig.~\ref{fig:figure_2}, the Brownian motion turns out to be the dominant noise source of the cantilever motion signal around the resonance frequency and it goes up to $S_X=15$ pm$^2$.Hz$^{-1}$ under vacuum ($P = 10^{-6}$ torr) and at room temperature. The oscillator is undergoing white noise of $S_F = 4 k_B T \Gamma$, that can be experimentally estimated at 13 fN/$\sqrt{\mathrm{Hz}}$ through $S_F (\omega_{res}) = (k_0/Q)^2 S_X(\omega_{res})$, where $Q$ is the quality factor. The probe cannot make out force gradient smaller than $\sqrt{S_{F'}} = \sqrt{S_F}/a_0$: in the present experiment, it therefore amounts to $1.3 \; 10^{-6}$ N.m$^{-1}/\sqrt{\mathrm{Hz}}$. Given a 0.33 Hz estimated bandwidth associated to the lock-in low pass, the theoretical measurement noise of 0.006 Pa is consistent with the 0.008 Pa standard deviation of data, which admit a Gaussian distribution. It generates a random error that is relevant at long range when compared to previously mentioned systematic error. As Casimir force decreases when moving away the two mirrors, the Brownian motion appears to set the upper boundary of the measurement range at 600 nm. This error analysis suggests that significant resolution improvements could be drawn, when working at low temperature. At 4 K, at least a factor of 10 of gain could be achieved in reducing thermal noise.

In conclusion, in this paper, emphasis is placed on analysing the main experimental issues associated to the realization of a Casimir microscope, aimed at comparing force behaviors between different sample configurations. Mechanical and frequency resonance drifts determine the main limitations of our present set-up. However, we have shown that methods used to correct these issues are efficient. Furthermore on the basis of the here presented experiences, a specific set-up used at liquid Helium temperature is currently designed in order to reduce the mechanical drift: at low temperature, this phenomenon will be very much reduced. At room temperature, current effort is applied towards implementing a distance separation optical sensor between the probe and the sample. Despite the severe constrain of no contact, which is here applied to the whole experimental process, we have shown that repeated quantitative Casimir force measurements can be carried out at different locations of the same sample with no irreversible change in the probe. A direct contact is not a prerequisite for Casimir force measurement; when combined with large X and Y displacement, this fact opens the way to quantitative and well characterized observations of new properties of fluctuation forces such as the Casimir force using structured surfaces at different scales. 

We are grateful to Serge Reynaud and Valery Nesvizhevsky for fruitful discussions. We thank Simon Le Denmat for sphere lever system preparation. Gold coating was performed in the Nanofab facilities (Institut N\'eel clean room).

\end{document}